%% file: msfp16-revised-correctionsinproof.tex
\documentclass{eptcs}
\providecommand{\event}{MSFP 2016}

\usepackage[utf8x]{inputenc}
\usepackage{alltt}
\usepackage{amsmath}
\usepackage{tikz}
\usepackage{tikz-qtree}
\usepackage{amssymb}
\usepackage[toc,page]{appendix}
\usepackage{bbm}

\DeclareUnicodeCharacter{8759}{$::$}
\DeclareUnicodeCharacter{8659}{$\Downarrow$}
\DeclareUnicodeCharacter{931}{$\mathtt{\Sigma}$}
\DeclareUnicodeCharacter{955}{$\mathtt{\lambda\!}$}

\usetikzlibrary{positioning}
\usetikzlibrary{calc}
\usetikzlibrary{arrows}
\usetikzlibrary{decorations.markings}

\newcommand*{\StrikeThruDistance}{0.15cm}%

\usetikzlibrary{matrix}

\tikzset{strike thru arrow/.style={
    decoration={markings, mark=at position 0.5 with {
        \draw [-] 
            ++ (-\StrikeThruDistance,-\StrikeThruDistance) 
            -- ( \StrikeThruDistance, \StrikeThruDistance);}
    },
    postaction={decorate},
}}

\tikzset{strike thru arrow left/.style={
    decoration={markings, mark=at position 0.5 with {
        \draw [-] 
            ++ (\StrikeThruDistance,-\StrikeThruDistance) 
            -- ( -\StrikeThruDistance, \StrikeThruDistance);}
    },
    postaction={decorate},
}}

\newcommand{\brl}{[}
\newcommand{\brr}{]}

\title{Variations on Noetherianness}
\author{
Denis Firsov \qquad\qquad Tarmo Uustalu \qquad \qquad Niccolò Veltri
\institute{Institute of Cybernetics at Tallinn University of Technology, Estonia}
\email{ \{denis,tarmo,niccolo\}@cs.ioc.ee }
}
\def\titlerunning{Variations on Noetherianness}
\def\authorrunning{D. Firsov, T. Uustalu \& N. Veltri}
\begin{document}

\maketitle

\begin{abstract}
  In constructive mathematics, several nonequivalent notions of
  finiteness exist. In this paper, we continue the study of Noetherian
  sets in the dependently typed setting of the Agda programming
  language. We want to say that a set is Noetherian, if, when we are
  shown elements from it one after another, we will sooner or later
  have seen some element twice. This idea can be made precise in a
  number of ways. We explore the properties and connections of some of
  the possible encodings. In particular, we show that certain
  implementations imply decidable equality while others do not, and we
  construct counterexamples in the latter case. Additionally, we
  explore the relation between Noetherianness and other notions of
  finiteness.
\end{abstract}

\section{Introduction}
\label{introduction}

To work with finite sets in the constructive setting of proof
assistants like Agda \cite{Agda}, which is the language we use in this
paper, we need to be able to say what a finite set is.  The
straightforward way of saying that a set is finite is to ask for an
enumeration of its elements together with the proof that the
enumeration is complete \cite{Kuratowski}. This idea can be formalized
as follows:
\begin{alltt}
Listable X = Σ[ xs ∈ List X ] ((x : X) → x ∈ xs)
\end{alltt}
(In Agda, \verb;Σ[ a ∈ A ] B a; is the type of dependent pairs of an
element \verb;a; of type \verb;A; and an element of type \verb;B a;. 
Note the unfortunate and confusing use of \verb;∈; instead of
\verb;:; for typing the bound variable in this notation.)

An alternative notion of finiteness found in the literature is
Noetherianness. Intuitively, a set \verb+X+ is Noetherian if, whenever we are shown enough
elements of \verb+X+, eventually we will have seen some element twice. Following Coquand and Spiwack \cite{Coquand}, this idea can be formalized as follows:
\begin{alltt}
data NoethAcc' (X : Set) (acc : List X) : Set where
  stop : Dup acc → NoethAcc' X acc
  ask  : ((x : X) → NoethAcc' X (x ∷ acc)) → NoethAcc' X acc
\end{alltt}
The auxiliary definition \verb;NoethAcc'; is parametrized by a set and
an accumulator list over this set. It has two constructors. The
constructor \verb;ask; says that, by constructing a proof of
\verb;NoethAcc' X (x ∷ acc); for all \verb+x : X+, one has 
constructed a proof of \verb;NoethAcc' X acc;. The constructor
\verb;stop; says that, if \verb;acc; already contains a duplicate, then
one gets a proof of \verb;NoethAcc' X acc;. Therefore, to construct a proof of
\verb;NoethAcc' X acc;, we must show that, if we ask for elements long
enough, then, independently of which elements we are presented, we can
eventually stop. Finally, we say that the set is Noetherian, if we can prove
\verb;NoethAcc'; starting with the empty accumulator:
\begin{alltt}
NoethAcc X = NoethAcc' X []
\end{alltt}

Let us prove that the set of Booleans is Noetherian. The proof is done by case
analysis and pattern matching under lambdas. Every time we ask for an
element, the proof tree branches according to whether the element provided is
\verb;true; or \verb;false;. Clearly, after asking for elements at least three
times, all branches can stop.
\begin{alltt}
BoolNoetherian : NoethAcc Bool
BoolNoetherian =
    ask (λ \{ true  → ask (λ \{ true  → stop dup0 
                             ; false → ask (λ \{ true  → stop dup1 
                                               ; false → stop dup2 \}) \}) 
           ; false → ask (λ \{ true → ask (λ \{ true  → stop dup2 
                                              ; false → stop dup1 \}) 
                             ; false → stop dup0 \})  \})
  where
    -- the implementation of dup* is left out for brevity
    dup0 : \{x : Bool\} → Dup (x ∷ x ∷ [])
    dup1 : \{x y : Bool\} → Dup (x ∷ y ∷ x ∷ [])
    dup2 : \{x y : Bool\} → Dup (x ∷ x ∷ y ∷ [])
\end{alltt}
The proof corresponds to the tree of the following shape:
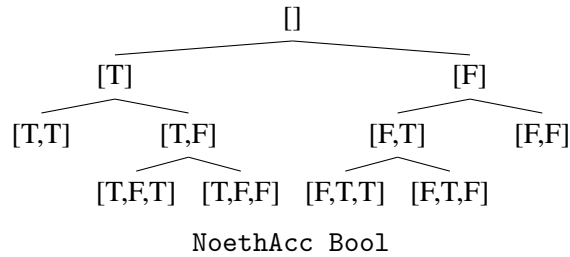
\begin{center}
    \begin{tabular}{c}
                \begin{tikzpicture}[level distance=22pt]
                \tikzset{frontier/.style={distance from root=90pt}}
                \Tree [.[] [ .[T]  [ .[T,T]  ] [ .[T,F] {\brl}T,F,T{\brr}  {\brl}T,F,F{\brr} ] ] [ .[F]  [ .[F,T]  {\brl}F,T,T{\brr}  {\brl}F,T,F{\brr} ] [ .[F,F] ] ]     ]
                \end{tikzpicture} \\
                \texttt{NoethAcc Bool} \\
    \end{tabular}
\end{center}
Note that the definition of \verb;NoethAcc; does not require us to terminate
immediately after discovering the first duplicate in the accumulator. Therefore, all
branches could be continued for some finite number of iterations.

We can already observe that Noetherianness has a number of interesting properties
in comparison with listability. A proof that a certain set is Noetherian does not provide
an access to elements of the set. This also implies that there is no information
about the size of the set, or even about inhabitedness of the set. Moreover, the proof objects
of Noetherianness are lightweight and do not induce excessive
computations. For example, consider the finite set \verb;Factors n;, which
contains all the factors of a natural number \verb;n;. It is straightforward to prove
that this set is both listable and Noetherian. However, the proof of listability can
cause the search of factors of \verb;n;, if pattern matching is performed on the
witness list. This causes problems for big values of \verb;n;.

In this paper, we continue the study of Noetherian sets and make the following
main contributions:
\begin{itemize}
\item In Section~\ref{different-encodings-of-noetherianness}, we provide a number of
implementations of the idea of Noetherianness and compare them to each other.

\item In Section~\ref{decidable-equality}, we show that the original definition
  of Noetherian sets by Coquand and Spiwack \cite{Coquand} (\verb;NoethAcc;) implies decidable
  equality on the set.

\item In Section~\ref{decidable-equality-for-noethaccs-a-counterexample}, we
  construct a class of finite sets with generally undecidable equality and prove that they are
  Noetherian for a particular variation of Noetherianness (\verb;NoethAccS;). This
  implies that some encodings are logically weaker than others.

\item In Section~\ref{connections-between-notions-of-finiteness}, we
  prove that the variations of Noetherianness introduced are
  nonequivalent by providing counterexamples. In addition, we establish
  connections with notions of listability, streamlessness, and
  almost-fullness. The relation between all the notions of finiteness
  presented in the paper is summarized in Figure~\ref{fig:rels}.
\end{itemize}
Section~\ref{related-work-and-conclusions} is devoted to related work,
conclusions, and further work. In Appendix~\ref{basic-definitions-in-agda}, we
give a detailed explanation of the basic Agda definitions we use (e.g., membership in a list,
duplicates, decidable equality, propositional types, etc). 

We used Agda 2.4.2.2 and Agda Standard Library 0.9 for this development. The
full Agda code of this paper can be found at
\url{http://cs.ioc.ee/~denis/noeth/}.

\section{Different Encodings of Noetherianness}
\label{different-encodings-of-noetherianness}

In the introduction, we have presented an encoding of Noetherianness,
called \verb+NoethAcc+. In order to construct a proof of
\verb+NoethAcc X+, we repeatedly ask for elements of \verb+X+ until we
end up with an element that we have seen twice. A couple of notes on this
encoding:
\begin{enumerate}
\item When we ask for an element of \verb+X+, we are given an
  \emph{arbitrary} one. In particular, we may receive an element
  that we have already seen and conclude using the constructor
  \verb+stop+. This may happen before having seen all the elements of
  \verb+X+.
\item A proof of \verb+NoethAcc X+ does not necessarily detect the first
  element that appears twice. We can keep asking for elements of
  \verb+X+ after we have already seen an element twice.
\end{enumerate}
The above observations expose the fact that there are some degrees
of freedom in the encoding of Noetherianness. This section is
devoted to the description of implementations of Noetherianness
alternative to \verb+NoethAcc+.

The first variation we present is called \verb+NoethAccS+. The extra ``S''
stands for ``strict''. In this implementation, we do not ask for arbitrary elements,
but only for \emph{fresh} ones. The idea can be formalized as follows:
\begin{alltt}
data NoethAccS' (X : Set) (acc : List X) : Set where
  ask : ((x : X) → ¬ x ∈ acc → NoethAccS' X (x ∷ acc)) → NoethAccS' X acc
\end{alltt}
The auxiliary definition \verb;NoethAccS'; has only one constructor
\verb;ask;. It says that, by constructing a proof of
\verb;NoethAccS' X (x ∷ acc); for all \verb+x : X+ that are not in the
accumulator \verb;acc;, one has constructed a proof of
\verb;NoethAccS' X acc;.  Then we say that the set is Noetherian, if we
can prove \verb;NoethAccS'; starting with empty accumulator:
\begin{alltt}
NoethAccS X = NoethAccS' X []
\end{alltt}
The base case is reached when, for all \verb;x : X;, we have that
\verb;¬ x ∈ acc; is false, i.e., can produce a proof of \verb;¬ ¬ x ∈ acc;.  Therefore, in
order to construct a proof of \verb+NoethAccS+, we repeatedly ask for fresh
elements of \verb;X; until no element can fail to be in the accumulator.

The predicate \verb;NoethAcc; is stronger than \verb;NoethAccS;, since every set
that satisfies \verb;NoethAcc; also satisfies \verb;NoethAccS;. More generally,
given a duplicate-free accumulator \verb;acc : List X;, we have that
\verb;NoethAcc' X acc; implies \verb;NoethAccS' X acc;. This fact is proved as
follows:
\begin{alltt}
NoethAcc'→NoethAccS' : \{X : Set\} → (acc : List X) → ¬ Dup acc
  → NoethAcc' X acc → NoethAccS' X acc
NoethAcc'→NoethAccS' acc ¬d (stop d) = ⊥-elim (¬d d)
NoethAcc'→NoethAccS' acc ¬d (ask n) =
  ask (λ x ¬m → NoethAcc'→NoethAccS' (x ∷ acc) (¬DupCons ¬d ¬m) (n x))
\end{alltt}
where \verb;⊥-elim; is the elimination principle of \verb;⊥; and \verb;¬DupCons;
constructs a proof of \verb;¬ Dup (x ∷ acc); from a proof of \verb;¬ Dup acc;
and a proof of \verb;¬ x ∈ acc;. The proof proceeds by induction on the proof of
\verb;NoethAcc' X acc;: if \verb;acc; contains a duplicate, then we can derive
\verb;⊥;, since by hypothesis \verb;acc; is duplicate-free; otherwise, we have a
proof \verb;n; of \verb;(x : X) → NoethAcc' X (x ∷ acc);. We ask for a fresh
element \verb+x+. This element can be added to the accumulator, which remains
duplicate-free, and we can invoke the induction hypothesis on 
\verb;n x : NoethAcc' X (x ∷ acc);.

The main statement follows by instantiating \verb;acc; with the empty list and
noting that the empty list is duplicate-free.
\begin{alltt}
NoethAcc→NoethAccS : \{X : Set\} → NoethAcc X → NoethAccS X
NoethAcc→NoethAccS n = NoethAcc'→NoethAccS' [] (λ ()) n
\end{alltt}
The converse implication does not generally hold, as we will demonstrate in
Section \ref{decidable-equality-for-noethaccs-a-counterexample}.

Alternatively, we can implement the idea behind \verb;NoethAccS; without an
explicit accumulator. In this variation, which we call \verb;NoethSet;, we also
ask for fresh elements of a given type \verb;X;, but instead of storing the
already seen elements in a list, we directly remove them from the set \verb;X;.
The predicate \verb;NoethSet; can be formalized as follows:
\begin{alltt}
data NoethSet (X : Set) : Set where
  ask : ((x : X) → NoethSet (X ∖ x)) → NoethSet X
\end{alltt}
where the type \verb;X ∖ x; of elements of \verb;X; different from
\verb;x; can be defined as follows:
\begin{alltt}
_∖_ : (X : Set) → X → Set
X ∖ x = Σ[ x' ∈ X ] ¬ x' ≡ x 
\end{alltt}
The type \verb;NoethSet; has only one constructor \verb;ask;.  It says
that, to construct a proof of \verb;NoethSet X;, one has 
to construct a proof of \verb;NoethSet (X ∖ x); for all
\verb+x : X+.
This encoding of Noetherianness was mentioned by Bezem et al.\cite{Uustalu}. 
Here the base case is reached when the type \verb+X+
becomes empty, i.e., from an inhabitant \verb+x : X+, we can derive
\verb+⊥+. Therefore, in order to construct a proof of \verb+NoethSet+,
we repeatedly ask for fresh elements of \verb+X+ and remove from
\verb+X+ the elements presented to us, until the set \verb+X+ becomes empty.

The encodings \verb+NoethAccS+ and \verb+NoethSet+ are logically
equivalent (\verb;NoethAccS→NoethSet; requires the
assumption of the principle of function extensionality).
\begin{alltt}
NoethAccS→NoethSet : \{X : Set\} → NoethAccS X → NoethSet X

NoethSet→NoethAccS : \{X : Set\} → NoethSet X → NoethAccS X
\end{alltt}
So \verb+NoethSet+ does not utilize an accumulator, but it still
remembers what elements of \verb+X+ have already been seen. These
elements correspond exactly to the ones that have been removed from
the set \verb+X+.

\subsection{The Noetherianness Game}
\label{the-noetherianness-game}

In this subsection, we give a game-theoretic description of Noetherianness. This
will help us develop other different variations on the theme. The Noetherianness
game, based on the encoding \verb;NoethAccS;, works as follows. Let \verb;X; be
a set. Two players participate in the game, the prover and the opponent. The
prover cannot see what is inside \verb;X; and repeatedly asks the opponent for
fresh elements of \verb;X;. The opponent knows exactly which elements are in
\verb;X; and answers the prover's queries by supplying a fresh element. The game
terminates, when the opponent cannot provide any fresh element to the prover. The
prover wins, if the game terminates in a finite number of steps, no matter what
the strategy of the opponent is. The opponent wins, if she can come up with a
strategy that makes the game non-terminating.

We present another variant on Noetherianness, also explainable along the lines of the
game-theoretical presentation given above. We call it \verb;NoethGame;. At each
turn the prover can not only ask for a fresh element of \verb;X; but also
provide an element of \verb;X;. Whenever the prover asks for a fresh element,
the opponent provides one. The game terminates only when the opponent cannot
satisfy the prover's request. The winning conditions are the same as in the
Noetherianness game. This idea can be formalized as follows:
\begin{alltt}
data NoethGame' (X : Set) (acc : List X) : Set where
  tell : (x : X) → NoethGame' X (x ∷ acc) → NoethGame' X acc
  ask  : ((x : X) → ¬ x ∈ acc → NoethGame' X (x ∷ acc)) → NoethGame' X acc
\end{alltt}
The constructor \verb;tell; says that having an inhabitant \verb;x : X; and a
proof of \verb;NoethGame' X (x ∷ acc); makes a proof of
\verb;NoethGame' X acc;. Finally, we define \verb;NoethGame X; as
\verb;NoethGame' X [];.
\begin{alltt}
NoethGame X = NoethGame' X []
\end{alltt}
The predicate \verb;NoethGame; is different in flavor from the Noetherianness
predicates introduced before. When constructing a proof of \verb;NoethGame X;, we
can not only ask for elements of \verb;X;, but also choose to provide elements of
\verb;X;, if we happen to know some.  The predicate \verb;NoethGame; is
particularly useful, when we have some kind of partial knowledge of the set
\verb;X;. Clearly, \verb;NoethAccS X; implies \verb;NoethGame X; by
construction.
\begin{alltt}
NoethAccS→NoethGame : \{X : Set\} → NoethAccS X → NoethGame X
\end{alltt}

We present another variant of Noetherianness, also based on the game-theoretic
intuition. We modify the rules of \verb;NoethGame;. At each step, the prover can
either ask for \emph{any} element of \verb;X;, provide an element of \verb;X;,
or win the game by giving a proof that the accumulator is exhaustive. Putting it
formally:
\begin{alltt} 
data NoethExpose' (X : Set) (acc : List X) : Set where
  stop : ((x : X) → x ∈ acc) → NoethExpose' X acc
  tell : (x : X) → NoethExpose' X (x ∷ acc) → NoethExpose' X acc
  ask  : ((x : X) → NoethExpose' X (x ∷ acc)) → NoethExpose' X acc

NoethExpose X = NoethExpose' X []
\end{alltt}
A set satisfying \verb;NoethExpose; has an interesting property. Once we know
an inhabitant of it, we know that the set is listable.

\begin{alltt}
NoethExpose→Listable : \{X : Set\} → X → NoethExpose X → Listable X
\end{alltt}
Given an element \verb;x : X;. A proof of \verb;NoethExpose X; is a tree that
we can walk up to a leaf by choosing the \verb;x;-th branch at each \verb;ask;
node. The \verb;tell; nodes construct a list and the leaf node of the path
contains a proof that the accumulator has all the elements of \verb;X;.

It is easy to see that every listable set satisfies
\verb;NoethExpose;. Moreover, \verb;NoethExpose X; implies \verb;NoethAcc X;. In
fact, just ask for an element of \verb;X;. At this point, the set \verb;X;
becomes listable. Hence we are done, since listable sets satisfy
\verb;NoethAcc;. The converse implication does not generally hold, as we will
demonstrate in Section \ref{separating-noethexpose-from-listability}.

\section{Decidable Equality}
\label{decidable-equality}

It has been shown \cite{Firsov-Uustalu} that every listable set has
decidable equality. In this section, we prove that the same holds for
Noetherian sets in the sense of \verb;NoethAcc;.
\begin{alltt}
NoethAcc→DecEq : \{X : Set\} → NoethAcc X → DecEq X
\end{alltt}
We give an informal account of the proof of \verb+NoethAcc→DecEq+. To
decide whether two elements \verb;x; and \verb;y; are equal, we proceed as
follows.
\begin{itemize}
\item
First, we repeatedly feed \verb;x; into the Noetherianness proof, until we get some
list \verb;xs; with the proof that it contains duplicates, say
\verb;d : Dup xs;. By construction, all elements of \verb;xs; are equal to
\verb;x;. The proof \verb;d; points to two different positions \verb;p1; and
\verb;p2; in \verb;xs;.
\item
Next, we repeat the procedure described above, this time feeding \verb;y;
instead of \verb+x+ at the \verb;p1;-th iteration. The procedure returns a proof
\verb;d' : Dup xs';. The list \verb;xs'; contains \verb+x+ at all positions
except for the \verb;p1;-th, where \verb;y; has been inserted. The proof
\verb;d'; points to two different positions \verb;p1'; and \verb;p2'; in
\verb;xs';.
\item
If \verb;p1; is equal to \verb;p1';, then clearly \verb;x; is equal to
\verb;y;. In the other case, if \verb;p1; is different from \verb;p1';, then also
\verb;x; differs from \verb;y;.
\end{itemize}

Notice that the same procedure cannot be replayed for
\verb+NoethAccS+, since we cannot feed the same element twice into a
proof of \verb+NoethAccS X+. 

As a corollary we obtain that, if a set satisfies \verb;NoethExpose;,
then it has decidable equality, since \verb;NoethExpose; is stronger
than \verb;NoethAcc;.

\subsection{A Counterexample to Decidable Equality for \texttt{NoethAccS}}
\label{decidable-equality-for-noethaccs-a-counterexample}

In this subsection, we assume the principle of function extensionality.
\begin{alltt}
funext : \{X Y : Set\} \{f g : X → Y\} → (∀ x → f x ≡ g x) → f ≡ g
\end{alltt}
The principle says that two functions are propositionally equal, if they deliver the same result for all arguments.  Assuming \verb;funext;, we show that it
is not the case that every set \verb;X; such that \verb;NoethAccS X; has
decidable equality. Note that this proves that not every set \verb;X; satisfying
\verb;NoethAccS X; satisfies \verb;NoethAcc X;.  Let us define a family
\verb;NotNotIn; of sets parametrized by a type \verb;X; and a list over
\verb;X;.
\begin{alltt}
NotNotIn : \{X : Set\} → List X → Set
NotNotIn \{X\} xs = Σ[ x ∈ X ] ¬ ¬ x ∈ xs
\end{alltt}
Since propositional equality is not decidable for a general type \verb;X; and
the functions of type \verb;¬ ¬ x ∈ xs; are all equal thanks to function
extensionality, we have that equality is not generally decidable for
\verb;NotNotIn xs;. Moreover, from the general decidable equality
on the type \verb;NotNotIn xs;, we can derive decidable equality for every type.
\begin{alltt}
EqNotNotIn→Eq : (\{X : Set\} (xs : List X) → DecEq (NotNotIn xs))
  → \{X : Set\} → DecEq X
\end{alltt}
The last step is to show that \verb;NotNotIn xs; satisfies
\verb;NoethAccS;.
\begin{alltt}
NoethAccSNotNotIn : \{X : Set\} → (xs : List X) → NoethAccS (NotNotIn xs)
\end{alltt}
The proof proceeds as follows. Apply \verb;length xs + 1; times the
constructor \verb;ask;. By doing so, we arrive at a list \verb;acc;
containing \verb;length xs + 1; different elements. However, the type
\verb;NotNotIn xs; has at most as many elements as there are positions
in the list \verb;xs;. From the two previous observations, we get a
contradiction, which corresponds to the base case of \verb;NoethAccS;. 

If we could derive \verb;NoethAcc; from \verb;NoethAccS;, then
\verb;NotNotIn xs; would have decidable equality by
\verb;NoethAcc→DecEq;. However, then \verb;EqNotNotIn→Eq; would allow us to
derive the decidable equality for every type, which is not plausible in a
constructive setting.

\section{Connections between Noetherianness and Other Notions of Finiteness}
\label{connections-between-notions-of-finiteness}

In the previous section, we have shown that the notions \verb;NoethAcc; and
\verb;NoethAccS; are different by constructing a ``separating'' class of sets, i.e., a class whose every member
satisfies \verb;NoethAccS;, but not \verb;NoethAcc; in general.  Given two notions of
finiteness \verb;F; and \verb;F';, we say that \verb;F; is separated from
\verb;F'; by a class of sets, if \verb;F' X; holds for all of its members, while \verb;F X; holding for all members implies some
nonconstructive principle. In this section, we show which other variations of
Noetherianness are separated. We also discuss the connection of Noetherian sets
with streamless sets and almost-full relations.

\subsection{Separating Listability from \texttt{NoethExpose}}
\label{separating-noethexpose-from-listability}

There exists a class of sets whose every member satisfies
\verb;NoethExpose; but is not listable. Consider any set \verb;X; such that
every two elements of \verb;X; are equal, i.e., a proposition. Then
\verb;NoethExpose X; holds, just ask for one element of \verb;X; and the element presented has already made the accumulator complete.
\begin{alltt}
NoethExposeProp : (X : Set) → isProp X → NoethExpose X
\end{alltt}
On the other hand, if we could construct a proof that any proposition \verb;X;
is listable, then, by checking whether the given list is empty or not, we could decide
the inhabitedness of \verb;X; (i.e., we could prove the law of excluded middle
for propositions).
\begin{alltt}
ListableProp→LEM : ((X : Set) → isProp X → Listable X)
  → (X : Set) → isProp X → X + ¬ X
\end{alltt}

\subsection{Separation from Bounded Sets}

A set \verb;X; is called bounded, if there exists a natural number (bound) \verb;n; such
that every list over \verb;X; with more than \verb;n; elements contains
duplicates.
\begin{alltt}
Bounded X = Σ[ n ∈ ℕ ] (xs : List X) → n ≤ length xs → Dup xs
\end{alltt}
Coquand and Spiwack \cite{Coquand} showed  that \verb;Listable X; implies
\verb;Bounded X;, and also \verb;Bounded X; implies
\verb;NoethAcc X;. Moreover, they proved that \verb;Listable; is
separated from \verb;Bounded; and most notably \verb;Bounded; is
separated from \verb;NoethAcc;. In fact, they proved that, if every
Noetherian set were bounded, then one could derive the limited principle of
omniscience (LPO).  The same class of sets separating \verb;Bounded; from
\verb;NoethAcc; also separates \verb;Bounded; from \verb;NoethExpose;.

We present a class of sets that separates \verb;NoethExpose; from \verb;Bounded;. Consider
a proposition \verb;X;. The set \verb;⊤ + X; (where \verb;⊤; is the unit type)
is bounded, since it contains at most two elements.
\begin{alltt}
MaybePropBounded : (X : Set) → isProp X → Bounded (⊤ + X)
\end{alltt}
On the other hand, if we could construct a proof that \verb;⊤ + X;
satisfies \verb;NoethExpose; for every proposition \verb;X;, then we
could derive the law of excluded middle for propositions. 
\begin{alltt}
MaybePropNoethExpose→LEM : (∀ X → isProp X → NoethExpose (⊤ + X))
  → ∀ X → isProp X → X + ¬ X
\end{alltt}
In fact, from the theorem \verb;NoethExpose→Listable;, we have that
\verb;NoethExpose (⊤ + X); implies that the set \verb;⊤ + X; is listable, since
\verb;⊤ + X; is inhabited for all \verb;X;. In turn, this implies that \verb;X;
is listable. But we already showed that, if every proposition is listable, then
we can derive the law of excluded middle for propositions
(\verb;ListableProp→LEM;).

Therefore \verb;NoethExpose; and \verb;Bounded; are separated from each other.
As a consequence, we have that \verb;NoethExpose; is separated from \verb;NoethAcc;.

\subsection{Streamless Sets}
A notion of finiteness similar to Noetherianness is the that of streamless
set. A set \verb;X; is streamless, if every stream (infinite list) over \verb;X;
contains duplicates.
\begin{alltt}
Streamless X = (xs : Stream X) → DupS xs
\end{alltt}
The formal definition of \verb;Stream; and \verb;DupS; can be found in
Appendix~\ref{basic-definitions-in-agda}. Coquand and Spiwack \cite{Coquand} showed
that every Noetherian set (in the sense of \verb;NoethAcc;) is
streamless.
\begin{alltt}
NoethAcc→Streamless : \{X : Set\} → NoethAcc X → Streamless X
\end{alltt}
Bezem et al.~\cite{Bezem} conjectured that it is unprovable that every
streamless set is Noetherian.  Parmann \cite{Parmann} also proved that
every streamless set has decidable equality under the hypothesis of function
extensionality (or stream extensionality, depending on the chosen representation
of streams).

The encoding of streamless sets admits variations similar to those for
Noetherianness introduced earlier. For example, we can define a set \verb;X; to be
streamless, if all duplicate-free colists (possibly infinite lists) have finite
length.
\begin{alltt}
StreamlessS X = (xs : Colist X) → ¬ DupC xs → xs ⇓ 
\end{alltt}
The formal definition of \verb;Colist;, \verb;DupC; and \verb;_⇓; can
be found in Appendix~\ref{basic-definitions-in-agda}.  Note that this strict
variation is similar to \verb;NoethAccS;. Moreover, \verb;NoethAccS;
is stronger than \verb;StreamlessS;.
\begin{alltt}
NoethAccS→StreamlessS : \{X : Set\} → NoethAccS X → StreamlessS X
\end{alltt}
As a corollary of \verb;NoethAccS→StreamlessS;, we have that in general
\verb;StreamlessS X; does not imply decidability of equality on
\verb;X;. Additionally, this shows that \verb;Streamless; and \verb;StreamlessS;
are separated similarly to \verb;NoethAcc; and \verb;NoethAccS;.

\subsection{Noetherian Sets and Almost-Full Relations}

Almost-full relations were introduced by Veldman and Bezem \cite{Veldman} for developing
an intuitionistic version of Ramsey theory. Vytiniotis et al.~\cite{Vytiniotis} analyzed almost-full relations in
connection to program termination and defined this concept 
as follows:
\begin{alltt}
data AF (X : Set) (R : X → X → Set) : Set where
  afzt  : ((x y : X) → R x y) → AF X R
  afsup : ((x : X) → AF X (λ y z → R y z + R x y)) → AF X R
\end{alltt}
A proof terminates, if the relation \verb;R; is total. Otherwise, we ask an
element \verb;x; from the opponent and we construct a bigger relation \verb;R';
such that, for all \verb;y z : X;, \verb;R' y z; if and only if \verb;R y z; or
\verb;R x y;. Then, by providing a proof of \verb;AF X R';, we conclude the proof
of \verb;AF X R;.

Vytiniotis et al.~\cite{Vytiniotis} remarked that the type \verb;AF X _≡_;
states that \verb;X; has finitely many inhabitants. We denote
\verb;AF X _≡_; by \verb;AFEq X;, and show that, for our hierarchy of encodings,
\verb;AFEq; is equivalent to \verb;NoethAcc;.
\begin{alltt}
AFEq X = AF X _≡_

AFEq→NoethAcc : \{X : Set\} → AFEq X → NoethAcc X

NoethAcc→AFEq : \{X : Set\} → NoethAcc X → AFEq X
\end{alltt}
Notice that the two above results cannot be proved by induction directly, since
the second constructor of \verb;AF; proceeds by growing the relation. Therefore,
we have to introduce a notion of Noetherianness for binary relations.
\begin{alltt}
data NoethAccR' (X : Set)(R : X → X → Set) (acc : List X) : Set where
  stop : DupR R acc → NoethAccR' X R acc
  ask  : ((x : X) → NoethAccR' X R (x ∷ acc)) → NoethAccR' X R acc

NoethAccR X R = NoethAccR' X R []
\end{alltt}
The notion \verb;NoethAccR'; is different from \verb;NoethAcc'; in the first
constructor, where instead of looking for duplicates in the accumulator we
search for related elements. This generalized notion \verb;NoethAccR; of Noetherianness for relations is equivalent to \verb;AF;.
\begin{alltt}
AF→NoethAccR : \{X : Set\}\{R : X → X → Set\} → AF X R → NoethAccR X R

NoethAccR→AF : \{X : Set\}\{R : X → X → Set\} → NoethAccR X R → AF X R
\end{alltt}
This equivalence can serve as an explanation of the rather unintuitive notion of
almost-fullness.

\section{Related Work and Conclusions}
\label{related-work-and-conclusions}

Finiteness in constructive mathematics has been studied by various authors
recently. Coquand and Spiwack \cite{Coquand} introduced four constructively
nonequivalent notions of finite sets in set theory \`a la Bishop: enumerated sets
(that we call listable sets), bounded size sets, Noetherian sets and streamless
sets. They showed how these different notions are connected and proved several
closure properties. Parmann \cite{Parmann} studied streamless sets in the setting of Martin-L\"of type theory. He showed that
streamless sets are closed under Cartesian product, if at least one of the sets
has decidable equality. Firsov and Uustalu \cite{Firsov-Uustalu} developed a practical toolbox for programming with listable subsets of base sets with decidable equality in Agda. Bezem et al.\ \cite{Uustalu} investigated a number of notions of finiteness of decidable subsets of natural numbers. 

\begin{figure}
\begin{center}
\input{diagram.tex}
\caption{Variations on Noetherianness.  LEM$_\mathrm{prop}$, LPO, and DEQ denote the law of excluded middle for propositions, the limited principle of omniscience, and the decidable equality for all types.}

\label{fig:rels}
\end{center}
\end{figure}
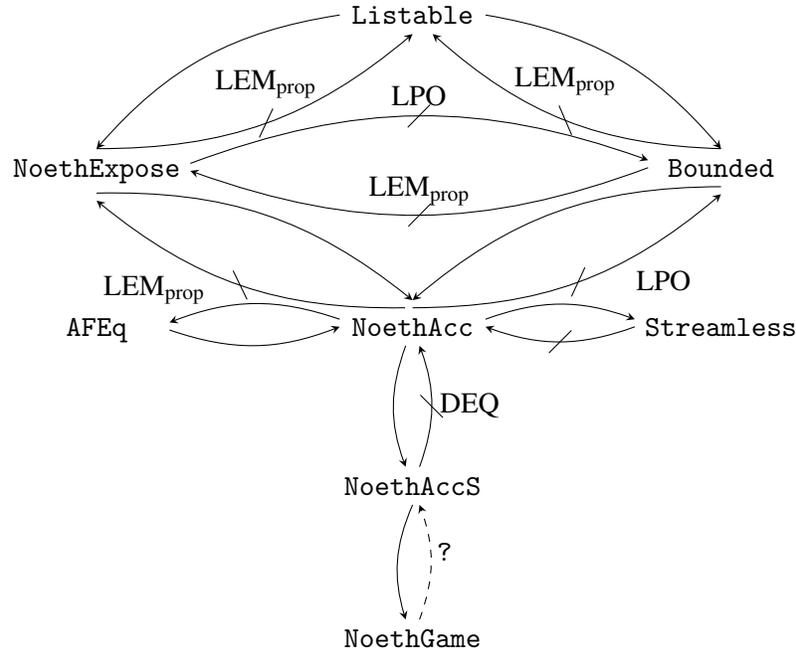

In this paper, we introduced several variations on the notion of Noetherian
set. Our current knowledge about the relations between different encodings is
summed up in Figure~\ref{fig:rels}.


Different encodings of Noetherianness all share the distinctive property of
hiding the elements of the set. Nonetheless some implementations ``reveal'' more
information about the set than others. We showed that \verb;NoethExpose;, and
most importantly \verb;NoethAcc;, allow one to construct a decider of equality for
the set, while such a decider cannot generally be built for sets satisfying
\verb;NoethAccS; or \verb;NoethGame;.

It remains open whether \verb;NoethAccS; and \verb;NoethGame; are equivalent notions
or not. A class of sets separating \verb;NoethAccS; from \verb;NoethGame; must
have the following properties: its members cannot have a computable bound on their size; they cannot have decidable equality.

Coquand and Spiwack \cite{Coquand} analyzed some closure properties of
\verb;NoethAcc;, such as closure under subsets, binary products and
coproducts. In this paper, we did not extend the study of such
properties for the variations on Noetherianness discussed.  This is a
possible direction for future work. Nevertheless, it is worth
mentioning that \verb;NoethAccS; is closed under quotients
(implemented as inductive-like types \`a la Hofmann \cite{Hofmann}) while
\verb;NoethAcc; is not.

In constructive mathematics, one encounters several further standard
ways of expressing finiteness, e.g., Dedekind finiteness. Clearly, one
can express finiteness in also in many exotic ways. We wonder whether
some form of unifying theory of useful notions finiteness in
constructive mathematics is possible.

\paragraph{Acknowledgement}
This research was supported by the Estonian Ministry of Education and Research
institutional research grant no.~IUT33-13, the Estonian Science Council personal research grant \linebreak no.~PUT763 and the Estonian Science Foundation
grant no.~9475.

\newcommand{\doi}[1]{\href{http://dx.doi.org/#1}{doi: #1}}

\bibliographystyle{eptcs}

\appendix

\section{Basic Definitions in Agda}
\label{basic-definitions-in-agda}

\paragraph{Membership in a list} We define the inductive type of proofs that an element
\verb;x; is in a list \verb;xs;.
\begin{alltt}
data  _∈_ \{X : Set\} (x : X) : List X → Set where
  here  : \{xs : List X\} → x ∈ x ∷ xs
  there : \{y : X\} \{xs : List X\} → x ∈ xs → x ∈ y ∷ xs
\end{alltt}
The constructor \verb;here; says that the head of a list is a member of
the list, and the constructor \verb;there; says that any element of the tail of a
list is also an element of the entire list.
\paragraph{Duplicates in a list} We define the type of proofs that the list contains duplicates:
\begin{alltt}
data Dup \{X : Set\} : List X → Set where
  duphere  : \{x : X\} \{xs : List X\} → x ∈ xs → Dup (x ∷ xs)
  dupthere : \{x : X\} \{xs : List X\} → Dup xs → Dup (x ∷ xs)
\end{alltt}
The list \verb;x ∷ xs; contains duplicates, if the element \verb;x; appears
in the tail \verb;xs; or if \verb;xs; contains duplicates.
\paragraph{Generalized types for membership and duplicates in a list} 
An inhabitant of the type \verb;x ∈ xs; is a proof that the list \verb;xs;
contains an element \emph{equal} to \verb;x;. This can generalized by
replacing  equality by an arbitary binary relation \verb;R;:
\begin{alltt}
data  MemR \{X : Set\}(R : X → X → Set) (x : X) : List X → Set where
  here  : \{y : X\} \{xs : List X\} → R y x → MemR R x (y ∷ xs)
  there : \{y : X\} \{xs : List X\} → MemR R x xs → MemR R x (y ∷ xs)
\end{alltt}
The type \verb;MemR R x xs; contains proofs that the element \verb;x; is related by \verb;R; to
some element in the list \verb;xs;. A similar generalization is possible for duplicates:
\begin{alltt}
data DupR \{X : Set\}(R : X → X → Set) : List X → Set where
  duphere  : \{x : X\} \{xs : List X\} → MemR R x xs → DupR R (x ∷ xs)
  dupthere : \{x : X\} \{xs : List X\} → DupR R xs → DupR R (x ∷ xs)
\end{alltt}
The type \verb;DupR R xs; contains proofs that the list \verb;xs; contains
a pair of elements related by \verb;R;.

\paragraph{Decidable equality}
If \verb;P; is a set, then \verb;Dec P; is a type of proofs of 
\verb;P; or not \verb;P;:
\begin{alltt}
data Dec (P : Set) : Set where
  yes : (prf :  P)  → Dec P
  no  : (prf : ¬ P) → Dec P
\end{alltt}
Here \verb;yes; and \verb;no; are two constructors of \verb;Dec P;.
The former takes a proof of \verb;P; as its argument while the latter
takes a proof of \verb;¬ P; (i.e., \verb;P → ⊥;).

Now, we say that \verb;X; has decidable equality, if, for any \verb;x₁;
and \verb;x₂; of type \verb;X;, we have \verb;Dec (x₁ ≡ x₂);:
\begin{alltt}
DecEq : Set → Set
DecEq X = (x₁ x₂ : X) → Dec (x₁ ≡ x₂)
\end{alltt}

\paragraph{Propositional types} 
We say that the type \verb;X; is a proposition, if any two elements of \verb;X; are
equal, i.e., if the type has at most one element:
\begin{alltt}
isProp : Set → Set
isProp X = (x₁ x₂ : X) → x ≡ x₂
\end{alltt}
For example, the empty and unit types are propositions.

\paragraph{Function extensionality}
Two functions are extensionally equal, if they return the same value
when applied to the same input. The principle of function
extensionality asserts that two functions are equal, if they are extensionally equal.
\begin{alltt}
funext : \{X Y : Set\} \{f g : X → Y\} → ((x : X) → f x ≡ g x) → f ≡ g
\end{alltt}
The principle of function extensionality is assumed in the proof of
\verb;NoethAccS→NoethSet; and in Section
\ref{decidable-equality-for-noethaccs-a-counterexample}. 

\paragraph{Membership and duplicates in a stream}
Streams are ``infinite lists'', defined coinductively as
follows:
\begin{alltt}
data Stream \{X : Set\} : Set where
  _∷_ : X → ∞ (Stream X) → Stream X
\end{alltt}
The membership relation and the predicate of duplicates 
are defined similarly to those for lists.
\begin{alltt}
data _∈S_ \{X : Set\} (x : X) : Stream X → Set where
  here  : \{xs : ∞ (Stream X)\} → x ∈S x ∷ xs
  there : \{y : X\} \{xs : ∞ (Stream X)\} → x ∈S ♭ xs → x ∈S y ∷ xs
\end{alltt}
\begin{alltt}
data DupS \{X : Set\} : Stream X → Set where
  duphere  : \{x : X\} \{xs : ∞ (Stream X)\} → x ∈S ♭ xs → DupS (x ∷ xs)
  dupthere : \{x : X\} \{xs : ∞ (Stream X)\} → DupS (♭ xs) → DupS (x ∷ xs)
\end{alltt}

\paragraph{Membership, duplicates and finite length for colists}
Colists are ``possibly infinite lists'', defined coinductively as
follows:
\begin{alltt}
data Colist \{X : Set\} : Set where
  []  : Colist X
  _∷_ : X → ∞ (Colist X) → Colist X
\end{alltt}
The membership relation and the predicate of duplicates 
are defined similarly to those for lists and streams.
\begin{alltt}
data _∈C_ \{X : Set\} (x : X) : Colist X → Set where
  here  : \{xs : ∞ (Colist X)\} → x ∈C x ∷ xs
  there : \{y : X\} \{xs : ∞ (Colist X)\} → x ∈C ♭ xs → x ∈C y ∷ xs
\end{alltt}
\begin{alltt}
data DupC \{X : Set\} : Colist X → Set where
  duphere  : \{x : X\} \{xs : ∞ (Colist X)\} → x ∈C ♭ xs → DupC (x ∷ xs)
  dupthere : \{x : X\} \{xs : ∞ (Colist X)\} → DupC (♭ xs) → DupC (x ∷ xs)
\end{alltt}
A colist has finite length, if it is a list after all. Formally, a colist has
finite length, if it satisfies the following inductively defined predicate:
\begin{alltt}
data _⇓  \{X : Set\} : Colist X → Set where
  []  : [] ⇓
  _∷_ : (x : X) \{xs : ∞ (Colist X)\} → (♭ xs) ⇓ → (x ∷ xs) ⇓
\end{alltt}

\end{document}

%% file: diagram.tex

\begin{tikzpicture}
  \matrix (m) [matrix of math nodes,row sep=4em,column sep=5em,minimum width=5em]
  {
      & \texttt{Listable} &   \\
     \texttt{NoethExpose} &  & \texttt{Bounded} \\
      \texttt{AFEq}& \texttt{NoethAcc}  & \texttt{Streamless}\\
      & \texttt{NoethAccS} & \\
      & \texttt{NoethGame} & \\};
  \path[-stealth]
    (m-1-2.west) edge[bend right=20] node [left] {} (m-2-1.north)
    (m-2-1.north) edge [strike thru arrow, bend right=20] node [above,yshift=0.2cm] {LEM$_\mathrm{prop}$}  (m-1-2.south)

    (m-1-2.east) edge[bend left=20] node [left] {} (m-2-3.north)
    (m-2-3.north) edge [strike thru arrow left, bend left=20] node [above,yshift=0.2cm] {LEM$_\mathrm{prop}$}  (m-1-2)

    ([yshift=0.1cm]m-2-1.east) edge[strike thru arrow, bend left=20] node [above] {LPO} ([yshift=0.1cm]m-2-3.west)    
    (m-2-3.west) edge[strike thru arrow, bend left=20] node [above] {LEM$_\mathrm{prop}$} (m-2-1.east)

    (m-2-3.south) edge[bend right=20] node [left] {} ([yshift=0.1cm]m-3-2.north)
    (m-3-2.north) edge [strike thru arrow, bend right=20] node [below,pos=0.8,yshift=-0.3cm] {LPO} ([yshift=-0.1cm]m-2-3.south)

    (m-2-1.south) edge[bend left=20] node [left] {} ([yshift=0.1cm]m-3-2.north)
    ([xshift=-0.1cm]m-3-2.north) edge [strike thru arrow left, bend left=20] node [below,pos=0.8,yshift=-0.3cm] {LEM$_\mathrm{prop}$} ([yshift=-0.1cm]m-2-1.south)

    (m-4-2) edge[strike thru arrow, bend right=20] node [right,pos=0.7,yshift=-0.3cm] {DEQ}  (m-3-2)
    (m-3-2) edge[bend right=20]  (m-4-2)

    (m-4-2.south) edge[bend right=20] node [left] {} (m-5-2)
    (m-5-2) edge [dashed, bend right=20] node [right,pos=0.8,yshift=-0.3cm] {\texttt{?}} (m-4-2)

    (m-3-1.east) edge[bend right=20]  (m-3-2.west)    
    ([yshift=0.1cm]m-3-2.west) edge[bend right=20]  ([yshift=0.1cm]m-3-1.east)    

    (m-3-3.west) edge[strike thru arrow, bend left=20]  (m-3-2.east)    
   ([yshift=0.1cm]m-3-2.east) edge[bend left=20]  ([yshift=0.1cm]m-3-3.west)    

    ;

\end{tikzpicture}
